# Entropy Production and Luminosity–Effective Temperature Relation for Main-Sequence Stars

**Leonid M. Martyushev [1,2,*] and Sergey N. Zubarev [1]**

[1] Ural Federal University, Mira St. 19, 620002 Ekaterinburg, Russia;
[2] Institute of Industrial Ecology, S Kovalevskoi St. 20a, 620219 Ekaterinburg, Russia
*leonidmartyushev@gmail.com

Based on the maximum entropy production principle, a relation between luminosity and effective temperature for main-sequence stars is obtained. Simplicity of the derivation and absence of any empirical parameters in the result is a fundamental difference of the present method from the classic ones where equations of stellar structure are analyzed. Using available photometric data (Webda, GCG) for more than 7.5 thousand stars, it is shown that the obtained luminosity–temperature relation is better than previously used ones.

## 1. Introduction

Main-sequence stars (MS) of the Hertzsprung–Russell diagram (HR) are the most common types of stars in the Universe. This is due to the fact that, for these stars, the main source of energy is the thermonuclear reaction of helium synthesis from hydrogen which takes approximately 90% of evolution time for the majority of stars. The main sequence lies in the neighborhood of the diagonal line in the HR diagram running from the upper-left corner (hot and bright stars) to the lower-right corner (cooler and less bright stars). For the sake of convenience, astrophysicists often convert the HR diagram quantities observed by astronomers (apparent stellar magnitude and color index) to those universally accepted in physics, i.e. luminosity ($L$) and effective temperature ($T$). Different calibrations are used for this conversion which represents a rather complex and approximate procedure [1-5]. This converted HR diagram is essential basic information for astrophysicists for verifying and correcting theoretical models of stars. As is known, the equations of stellar structure are extremely complex and non-linear [6-9]. They additionally include semi-empirical relations and parameters. As a consequence, even when analyzing MS stars that are relatively simple from a theoretical standpoint (specifically, for obtaining a relation of $L$ and $T$), numerical methods are applied, which leads to additional errors and ambiguity. Obviously, all this is very inconvenient for theoretical consideration. As a result, various approaches are developed in order to approximately solve (in particular, using dimensional analysis) the equations of stellar structure [6-9]. Ultimately, the following relation is obtained:

$$\text{Log } L = A \text{ Log} T + B, \qquad (1)$$

where the parameter A depends on the type of fusion reaction, the law of energy release, and the opacity law. Different authors obtain different values for this parameter depending on their approximations. Thus, in the case of a pp reaction depending on the opacity law (Kramer's or Thomson scattering), $A$ equals 4.1 or 5.6 [6]; 4.2 or 4.3 [7], respectively. In [8] $A$ = 5.6 or 6 (Thomson scattering, depending on a power value in the law of energy release); and in [9], $A$ = 5.6 (Thomson scattering). In the case of a CNO reaction depending on the opacity law (Kramer's or Thomson scattering), $A$ equals 5.5 or 8.6 [6]; 6 or 5.4 [7], respectively. In [8] and [9], for a CNO reaction, coefficients for Thomson scattering only are given: A= 8.3 or 8.7 (depending on a power value in the law of energy release) [8] and A = 8.4 [9]. The parameter $B$ in Eq. (1) is, in fact, chosen empirically.

The drawback of Eq. (1) is that multiple assumptions and approximations are required to derive it. Particularly, the laws of opacity and energy generation are approximately represented by power laws, and each approximation is valid in some range of temperature and density. All this makes it difficult to draw a quantitative comparison between Eq.(1) and relations of $L$ to $T$ obtained from astronomical data for real stars. These problems obviously result from the fact that the equations of the stellar structure are a too detailed level of description (both temporally and spatially) for deriving such integral quantities as $L$ and $T$ associated with a star as a whole. A question arises: is it possible to obtain information about the important relation $L(T)$ for MS stars using a method different from the one developed about a hundred years ago and associated with the names of such classics as Eddington, Schwarzschild, etc.? The purpose hereof is to demonstrate that this can be very easily done using general laws of nonequilibrium thermodynamics as a basis. At the same time, the obtained relation will be free from any empirically-chosen parameters and will describe available experimental data no worse than the known approximations.

The procedure is as follows. Firstly, we derive the relation of $L(T)$. Then, using photometric data for MS stars from the Webda and GCG databases as well as up-to-date calibration, we obtain a rather large volume of luminosity and temperature data covering a wide range of values. Finally, taking into account data errors, we compare the resulting data with the available analytical relations of $L$ to $T$ and draw conclusions.

## 2. Entropy production of a star and derivation of $L(T)$

A star continuously radiates a large amount of energy into outer space as a result of thermonuclear reactions. From this viewpoint, it represents an irreversible and open system. Classical nonequilibrium thermodynamics is a traditional approach to analyze such systems [10-12]. This method has a great generality but at the same time often omits various secondary details of occurring processes. Let us base our approach on nonequilibrium thermodynamics.

For MS stars, an approximation of the local thermodynamic equilibrium and of the stationarity of internal nonequilibrium processes is traditionally used for the analysis [6-9]. Further, a description of the outer envelope of a MS star (photosphere) as a black body having an effective temperature and radiating in accordance with the Stefan–Boltzmann law is a universally accepted and rather good approximation [6-9]. We need these assumptions too.

The entropy production (or entropy increment per unit of time resulting from occurring irreversible processes) $\Sigma$ is a basic quantity of nonequilibrium thermodynamics. Entropy production provides integral data about a star's irreversible processes in the most concise form. For a stationary irreversible process, the entropy production of a star equals the entropy flux from its surface. In the case of the accepted approximation of the black body, this flux is well known from the times of M. Planck [13]. As a result, the entropy production of a star has the form:

$$\Sigma = \frac{4}{3}\frac{L}{T}. \qquad (2)$$

As is seen, the entropy production of a star has a very simple form and can be found using available astronomical experimental data. While entropy production is most crucial for nonequilibrium thermodynamics having almost a hundred-year history, this quantity was very rarely employed in theoretical studies of stars. We are aware of just three groups of papers (one of which represents our own recent publications) which only calculate and analyze this quantity for a number of stars [14-19]. This lack of scrutiny is apparently due to the fact that, at first glance, it is not possible to obtain any new important information from this integral parameter for astrophysics which is traditionally based on solving a rather detailed and complete set of equations of stellar structure. We shall demonstrate here that such an opinion is erroneous.

According to one of the basic variational principles of nonequilibrium thermodynamics, the maximum entropy production principle (MEPP), a nonequilibrium system evolves in such a way as to maximize, under restrictions present in the system, its local entropy production (or entropy production density) [20-24]. As a consequence, if there are multiple systems in a nonequilibrium medium, the system with maximum local entropy productions will be the most viable during evolution. Such systems can co-exist over a large time span only when their local entropy productions are equal.

The averaged local (per unit volume) entropy production $\Sigma_V$ of a ball-shaped star (with its radius expressed using the Stefan–Boltzmann formula through $L$ and $T$) can be easily derived from Eq. (2) (see, e.g., [18,19]):

$$\Sigma_V = \chi \cdot T^5/\sqrt{L}, \qquad (3)$$

where $\chi = 8 \cdot \sigma^{3/2} \cdot \pi^{1/2}$, $\sigma$ is the Stefan–Boltzmann constant.

According to the formulated principle, we shall assume that the quantity $\Sigma_V$ is the same for all MS stars. Indeed, these stars were formed in some region of a galaxy due to a number of nonequilibrium processes and then they co-exist for the most part of their lives. Our previous calculations confirm this assumption, see Fig. 1 [18, 19]. As is seen, $\Sigma_V$ remains constant for different MS stars when luminosity changes by approximately six orders of magnitude (from 0.002 to 6000 $L / L_\odot$). For comparison, the specific entropy production per unit mass is given. The Sun is the most studied star whose thermodynamic parameters were measured with the greatest precision. So, based on the above, we shall assume that the specific entropy production of any MS star is the same as that of the Sun $\Sigma_{V\odot}$:

$$\Sigma_{V\odot} = \chi \cdot T_\odot^5/\sqrt{L_\odot}, \qquad (4)$$

where $T_\odot$ and $L_\odot$ are the Sun's effective temperature and luminosity, respectively.

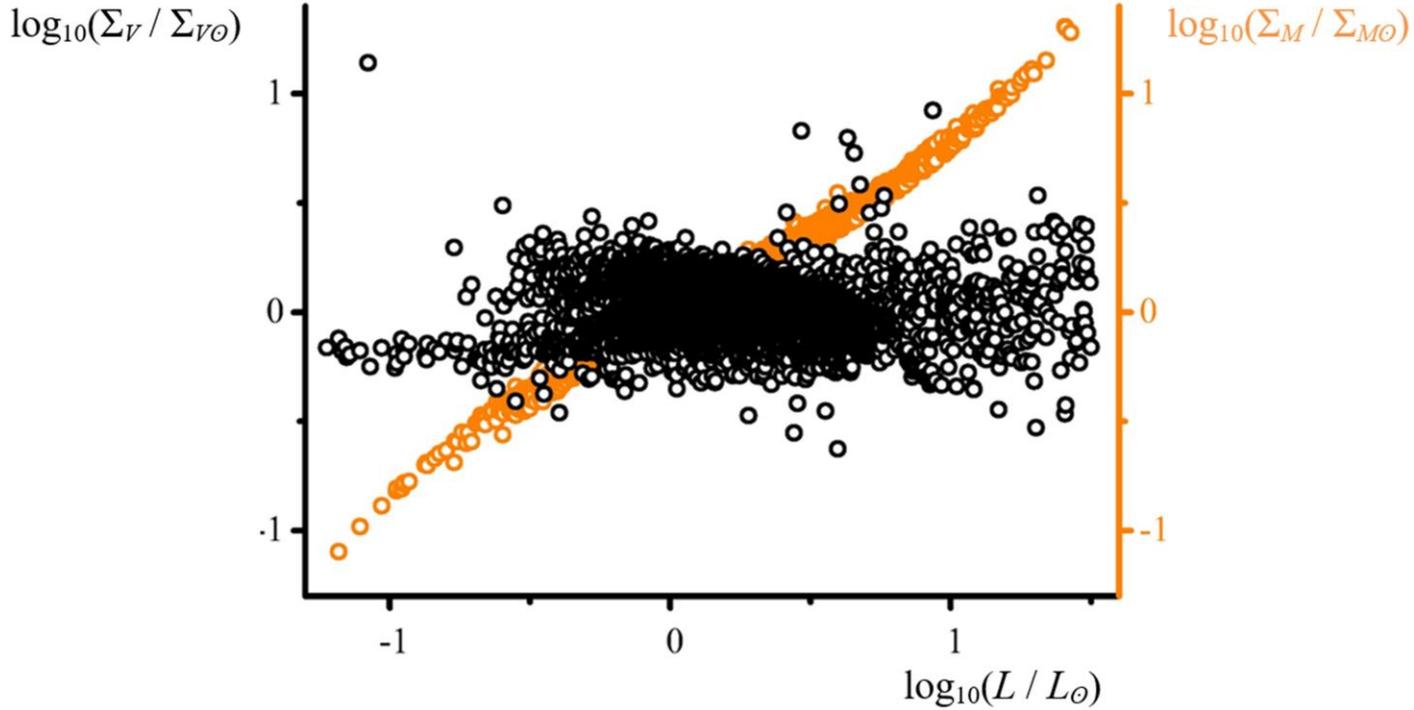

Fig. 1. Relation of the local entropy production per unit volume $\Sigma_V$ and the local entropy production per unit mass $\Sigma_M$ to the luminosity $L$ (see Appendices for more detailed information). Star masses are calculated using [25] whose semi-empirical relation is applicable to FGK stars. The data here and below are normalized to solar magnitudes. Their values are as follows: $T_\odot = (57 \pm 1.2) \times 10^2$ K, $L_\odot = (3.8 \pm 0.6) \times 10^{26}$ W, $\Sigma_{V\odot} = (6 \pm 2) \times 10^{-5}$ W K$^{-1}$ m$^{-3}$, $\Sigma_{M\odot} = (4.4 \pm 0.7) \times 10^{-8}$ W K$^{-1}$ kg$^{-1}$.

As a result, according to Eq.(3) and Eq.(4):

$$\text{Log } L = 10\text{Log } T + 2\text{Log}(\chi / \Sigma_{V\odot}), \qquad (5)$$

or

$$\mathrm{Log}(L/L_\odot) = 10\mathrm{Log}(T/T_\odot). \tag{6}$$

Thus, proceeding from the most general thermodynamic ideas, we manage to obtain the $L(T)$ relation whose form matches that of the commonly accepted one as represented by Eq. (1). It is an important feature of the found relation that the derived law has no empirically-chosen parameters and presupposes a single inclination angle of the $L(T)$ line (for logarithmic plotting) as opposed to several possible coefficients $A$ in Eq. (1). How does the proposed relation (6) describe available experimental data? Before answering this question, we shall produce HR diagrams in the coordinates $L$ and $T$ for MS stars for a number of star clusters.

### 3. Producing relations of $L$ to $T$ from photometric data.

Photometric data from the Webda [26] and GCG [27] databases (for MS stars belonging open and globular clusters, respectively) have been used. For every star of a cluster, Webda and GCG databases contain experimentally-obtained values of the apparent stellar magnitude $V$, the color index $(B–V)$, the color excess $E(B–V)$, the distance modulus $(m–M)_V$, etc. It is particularly reasonable and informative to analyze a sample of stars belonging to one cluster as their values of $E(B–V)$ and $(m–M)_V$ are relatively easy to find and are approximately the same – they are needed to calculate the normal color index of a star and its absolute magnitude. They have a standard and well-known calculation algorithm [18,19]. In order to obtain effective temperature from the photometry, a semi-empirical calibration proposed in Ref. [4] has been employed. An advantage of this calibration is that it is applicable to a very wide temperature range for the stars under study, from thousands to tens of thousands of degrees. A disadvantage is that metallicity [Fe/H] is not taken into account which may affect the calibration's accuracy. In terms of the accuracy of $T$ calculation, we have compared the calibration [4] with one of the most accurate semi-empirical calibrations, [3], that takes into consideration [Fe/H]. Unfortunately, the latter calibration is only applicable to FGK-class stars (which temperatures lie within 4,000 K to 7,500 K). It has been found that, for the absolute majority (98%) of stars to which the calibration [3] is applicable, the difference between temperatures calculated using the calibrations [3,4] does not exceed 6%. Thus, considering the small error, the semi-empirical calibration [4] has been employed for all calculations as being the most universal and applicable to the overwhelming majority of all MS stars. Additionally, the paper [4] proposes a semi-empirical relation to calculate bolometric correction which is needed to obtain absolute bolometric magnitude and, consequently, to calculate $L$ [18,19].

$T$ and $L$ have been calculated for MS stars of 13 open clusters (NGC 884, NGC 869, IC 4725, NGC 2516, NGC 1039, NGC 3532, NGC 2099, NGC 2281, NGC 2506, NGC 2682, NGC 188, NGC 2632, Hyades) [26] and 2 globular ones (NGC 6121 и NGC 6656) [27]. Ages of the studied clusters, are from 7.1 to 10.1 log(year). Only stars with an over 90% probability of cluster membership and an RMS error for the B and V magnitude of less than 0.02 mag have been considered. Calculations have included 7,636 stars in total. Calculation results for $T$ and $L$ grouped by clusters are given in Appendices. For illustration, a number of calculated HR diagrams for MS stars are shown in Fig. 2.

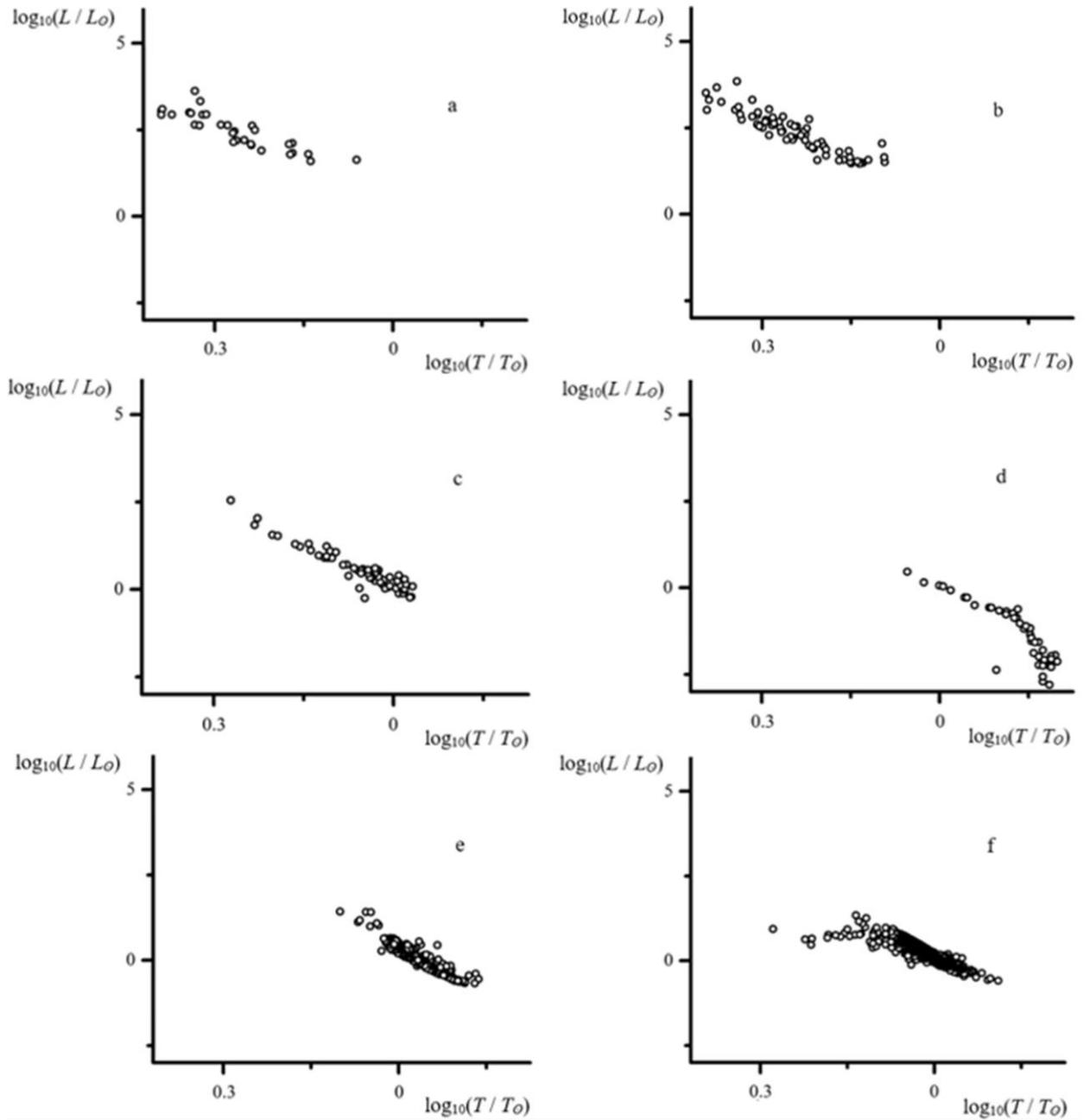

Fig. 2. HR diagrams for MS stars with the coordinates $L - T$ for several clusters. (a) NGC 884 (7.10), (b) NGC 869 (7.28), (c) NGC 1039 (8.42), (d) Hyades (8.90), (e) NGC 188 (9.80), (f) NGC 6121 (10.10). Cluster ages are specified in brackets using the form: $\log_{10}$ (t, year).

Based on the accuracy of the photometric data and the error of the semi-empirical calibration, an error of calculating thermophysical quantities for MS stars was obtained. The accuracy of photometric data represents the main contribution to the calculation error. It has been found that, for stars with temperatures below 4,000 K, the error of $T$ and $L$ does not exceed 4% and 44%, respectively; for stars with temperatures from 4,000 K to 10,000 K, the error does not exceed 8% and 29%, respectively; for stars with temperatures from 10,000 K to 14,500 K, the error does not exceed 20% and 66%, respectively; and for stars with temperatures over 14,500 K, the error exceeds 20% and 66%, respectively. We shall note that the mentioned error results in a major error when calculating specific entropy production based on Eq. (3) (see the scatter of points in Fig. 1). So, for $\Sigma_V$, the error does not exceed 38% for stars with temperatures below 4,000 K; 43% for stars with temperatures from 4,000 K to 10,000 K; 91% for stars with temperatures from 10,000 K to 14,500 K, and exceeds 91% for stars with temperatures over 14,500 K. Consequently, we have not considered or analyzed stars hotter than 14,500 K.

### 4. Comparison of theoretical $L(T)$ relations and the relations found using photometric data

The values of $L$ and $T$ found using the photometric data have, as is shown above, a rather significant error. Further, due to different levels of knowledge about different star clusters, their ages and distances, these clusters have considerably different ranges of $L$ and $T$ values with sufficient data (see Fig. 2). It is therefore reasonable to present all the obtained data (for 7,636 MS stars) in one final HR diagram. Certainly, such a representation form is not always correct. So, as is known, the parameter [Fe/H] affects positioning of MS stars in an HR diagram. However, this influence is weak and difficult to detect using traditional universal calibrations [1-4] and experimental photometric data (with their errors) available for the vast majority of stars. Fig. 3 and Fig. 4 show a combined HR diagram and its various analytical approximations. As follows from the comparison between Fig. 3 and Fig. 4, the relation Eq. (6) obtained herein describes the experimental points no worse than the traditional Eq. (1). Moreover, if the entire studied range of $L(T)$ values is considered while taking into account high-temperature and especially low-temperature stars, Eq. (6) undoubtedly represents a more preferable approximation.

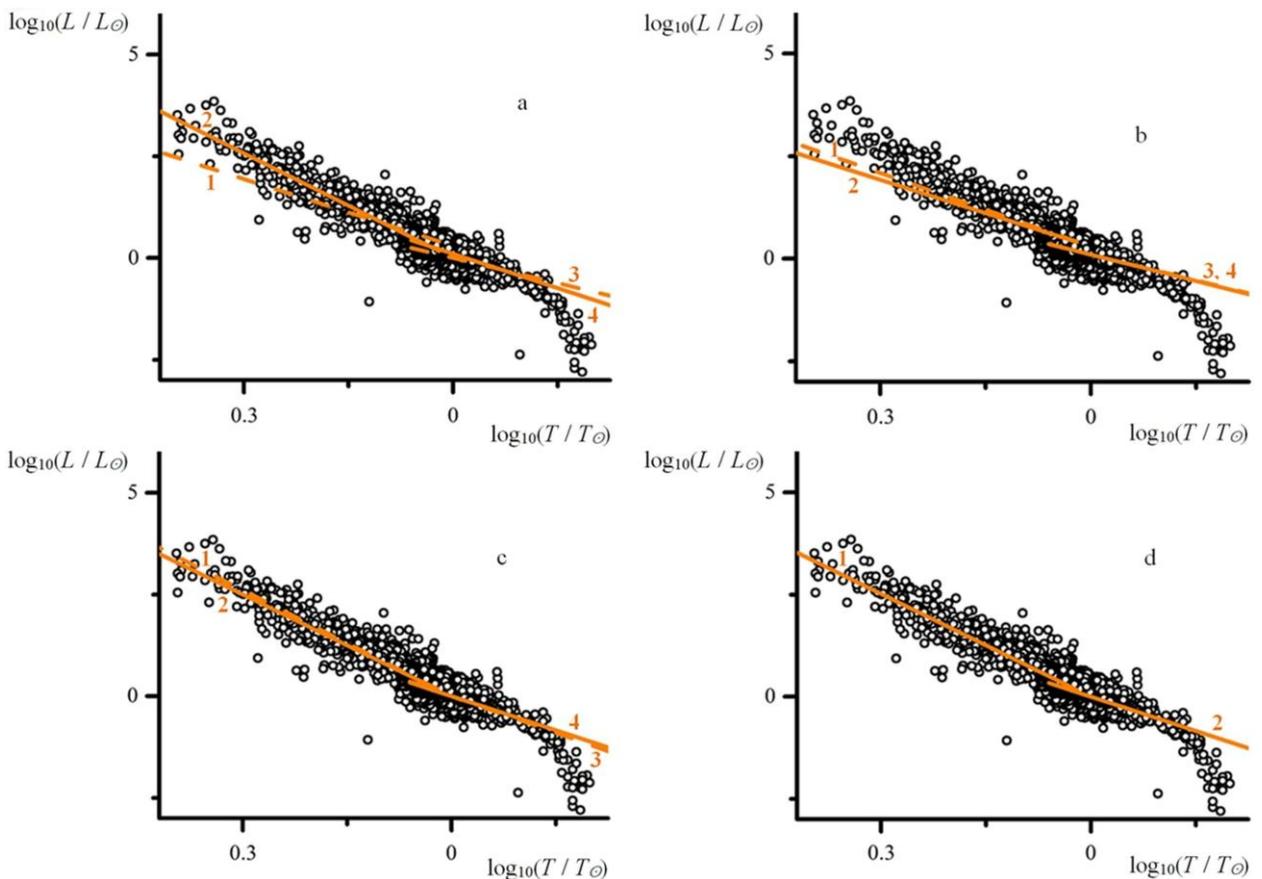

Fig. 3. The combined HR diagram based on the processed photometric data, and the theoretically-predicted $L(T)$ relations, Eq. (1).
  (a) A=5.5, B=0.3 (line 1); A=8.6, B=0 (line 2); A=4.1, B=0 (line 3); A=5.6, B=0.1 (line 4) [6].
  (b) A=6.0, B=0.3 (line 1); A=5.4, B=0.3 (line 2); A=4.2, B=0.1 (line 3); A=4.3, B=0.1 (line 4) [7].
  (c) A=8.7, B=0 (line 1); A=8.3, B=0 (line 2); A=6, B=0 (line 3); A=5.6, B=0 (line 4) [8].
  (d) A=8.4, B=0 (line 1); A=5.6, B=0 (line 2) [9].

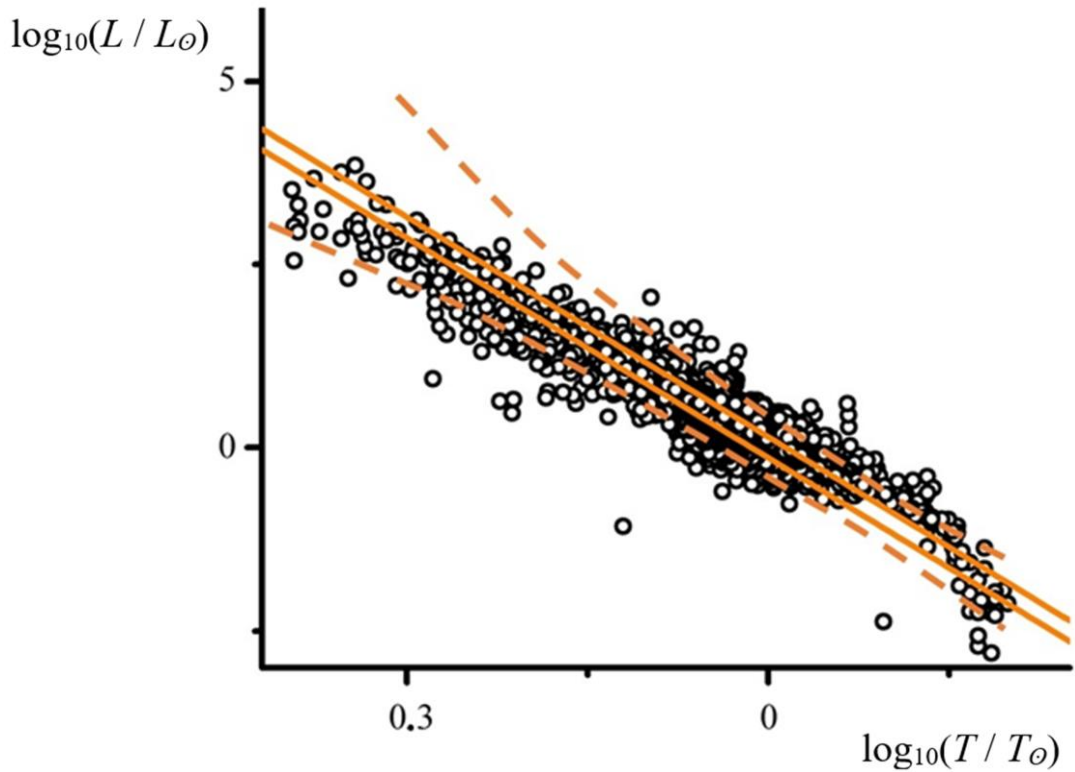

Fig. 4. The combined HR diagram based on the processed photometric data, and the *L(T)* relation predicted theoretically using MEPP, Eq. (6). The corridor shown by the solid line represents an *L(T)* region where such values of *L* and *T* may be found which are typical for stars with entropy productions matching that of the Sun. According to [28-36], $T_\odot = (57 \pm 1.2) \times 10^2$ K, $L_\odot = (3.8 \pm 0.6) \times 10^{26}$ W, and so, based on Eq. (6), the corridor's width is determined from the relation $\text{Log}(L / L_\odot) = 10\text{Log}(T / T_\odot) \pm (T_\odot \Delta L_\odot \pm L_\odot \Delta T_\odot)/ L_\odot T_\odot$, $\Delta L_\odot = 0.6 \times 10^{26}$ W, $\Delta T_\odot = 1.2 \times 10^2$ K. The dashed line shows the region in the HR diagram where, considering the current error in the accuracies of photometric data and calibration, experimental points are described by the derived Eq. (6).

## 5. Conclusion

The present study obtains, on the basis of MEPP, a new theoretical approximation of the important relation for MS stars: luminosity–effective temperature. Almost complete avoidance of assumptions typical for the traditional derivation based on consideration of a set of the equations of stellar structure represents a fundamental distinction of the provided derivation. The produced relation is also free from any adjustable or semi-empirical parameters. The obtained law describes existing experimental data no worse the known classic approximations (and even better, for extreme values in the *L(T)* diagram).

This approach is not to be considered as an alternative to the classic one. Apparently, the non-equilibrium thermodynamic method does not allow analyzing a star's temporal evolution at a sufficiently "small" space–time scale as it is a more rough, integral method. If we draw an analogy with mechanics, the given approach may be compared to a conservation law-based approach, whereas the traditional approach (solving a set of the equations of solar structure) may be considered similar to an approach where the fundamental equation of Newtonian dynamics is solved. As is known, for a number of problems, a solution involving the conservation law is much easier than with Newtonian dynamics; these methods, however, are not contrasted but complement each other well.

The fruitfulness of nonequilibrium thermodynamics demonstrated herein while addressing an important problem of the astrophysics of MS stars can apparently have other manifestations. In the future, it would be interesting to apply the same approach to calculate parameters for other types of stars as well as to analyze temporal evolution of stars from their formation to death.